\documentclass[american,aps,preprint]{elsarticle}
\usepackage[T1]{fontenc}
\usepackage[latin9]{inputenc}
\usepackage{amstext}
\usepackage{graphicx}
\usepackage{babel}
\begin{document}

\title{Black hole complementarity: the inside view}

\author{David A. Lowe}

\ead{lowe@brown.edu}

\address{Department of Physics, Brown University, Providence, RI 02912, USA}

\author{Larus Thorlacius}

\ead{larus@nordita.org}

\address{Nordita, KTH Royal Institute of Technology and Stockholm University,
Roslagstullsbacken 23, SE-106 91 Stockholm, Sweden\\{\rm and}\\ University
of Iceland, Science Institute, Dunhaga 3, IS-107 Reykjavik, Iceland}
\begin{abstract}
Within the framework of black hole complementarity, a proposal is
made for an approximate interior effective field theory description.
For generic correlators of local operators on generic black hole states,
it agrees with the exact exterior description in a region of overlapping
validity, up to corrections that are too small to be measured by typical
infalling observers. 
\end{abstract}
\maketitle

\section{Introduction}

Black hole complementarity posits that a unitary and local description
of physics exists outside a stretched horizon, a timelike surface
a short distance from the event horizon of a black hole. The postulates
of \citep{Susskind:1993if} leave open the question of how to describe
the physics inside the horizon but based on the equivalence principle
it is reasonable to expect that a freely falling observer experiences
nothing out of the ordinary when crossing the horizon of a sufficiently
large black hole. If this expectation is indeed borne out, it also
seems reasonable that observations made inside a laboratory that enters
a black hole in free fall should be described, to within achievable
experimental precision, by a more or less conventional effective field
theory. It was already observed in \citep{Susskind:1993if} that this
effective description cannot be a local quantum field theory that
is simultaneously valid for distant observers and observers who have
entered the black hole in free fall. The problems that arise when
one attempts to implement unitary black hole evolution from the point
of view of distant observers in the context of a local effective field
theory that extends into the black hole interior were stated more
sharply in \citep{Lowe:2006xm}, where it was pointed out that observations
made on the outgoing Hawking radiation would project the quantum state
of the black hole and in effect burn up the inside observer. In fact,
no explicit measurements are needed - the effect follows from decoherence
due to the local coupling between the Hawking radiation and degrees
of freedom far from the black hole. More recently similar conclusions
were reached in \citep{braunstein,Almheiri:2012rt} by considering
the entanglement between outgoing Hawking modes at different times
during the evaporation. An alternative conclusion is that there is
no firewall but that the problem lies with applying local effective
field theory across the horizon \citep{Lowe:2006xm,LarjoLowe}. 

In the present work we construct an approximate effective field theory
for an observer who passes through the black hole horizon in free
fall. The construction follows up on our recent work in \citep{Lowe:2013zxa}
where the evolution of a black hole formed in a generic pure state
was considered and it was argued that a typical infalling observer
would not see any drama on their way towards the stretched horizon.
While this is a satisfying conclusion it does not answer the key question
of what happens to such an observer in the interior region, which
we take to include both the black hole region inside the event horizon
and the region between the event horizon and the stretched horizon.
In order to address that question we need to have a model for the
interior quantum evolution and the answer turns out to depend on the
model. If we, for instance, choose to use a local quantum field theory
on a set of time slices that cover the exterior region during much
of the black hole lifetime and also extend smoothly into the black
hole region, staying away from the strong curvature near the black
hole singularity, then we would conclude that either there is no information
about the black hole state carried in the Hawking radiation, as was
indeed concluded by Hawking \citep{Hawking:1976ra}, or that the equivalence
principle is violated, as was concluded by the authors of \citep{braunstein,Almheiri:2012rt}.
Our construction gets around this by patching together effective field
theories on either side of the stretched horizon in such a way that
a typical infalling observer will not see any drama until near the
black hole singularity. A prescription for the interior initial data
is provided which is formally nonunitary, but we argue this nonunitarity
is unobservable, and akin to the harmless nonunitarity introduced
by a finite proper distance cutoff in effective field theory around
an expanding cosmological background. This nonunitary step in constructing
an effective field theory description for an infalling observer does
not affect the unitarity of the evaporation process from the exterior
viewpoint, and is perhaps the key new element that allows us to evade
the arguments of \citep{braunstein,Almheiri:2012rt}. 

The construction only applies to a restricted class of observers and
it is restricted to a set of time slices that only cover a relatively
short period of time before and after the observer enters the black
hole. Our main claim is that, even with these restrictions imposed,
the resulting effective field theory can describe observations made
by a typical infalling observer to sufficient accuracy to conclude
that no drama is encountered until deep inside the black hole. 

An alternative approach to describing the interior physics, inspired
by the non-locality of string field theory \citep{Lowe:1995ac}, is
to look for a non-local formulation of quantum field theory on a continuous
background geometry. For recent work along those lines see \citep{Giddings:2013kcj,Giddings:2013noa}.
Another approach is that of fuzzball complementarity \citep{Mathur:2012jk,Mathur:2013gua}
which uses string theory degrees of freedom to build an interior description.
Fuzzball complementarity shares some features of our effective field
theory construction but there are important differences which we comment
on at the end of Section \ref{sec:Pull-Back/Push-Forward-Revisited}
below.

\section{Black hole geometry and infalling observers\label{sec:Black-hole-geometry}}

A black hole of mass $M$ formed in the gravitational collapse of
non-rotating neutral matter in $3+1$ dimensional asymptotically flat
spacetime will settle down to a metastable state in a time of order
$M$ as measured by distant observers and then slowly evaporates due
to Hawking emission in a time of order $M^{3}$. During the evaporation,
on time scales that are short compared to the black hole lifetime,
the geometry is well approximated by the static Schwarzschild solution

\[
ds^{2}=-\frac{32M^{3}}{r}e^{-\frac{r}{2M}}dUdV+r^{2}d\Omega^{2}
\]
written here in Kruskal coordinates, related to the familiar Schwarzschild
coordinates $r,t$ by 
\begin{eqnarray*}
V & = & \left(1-\frac{r}{2M}\right)^{1/2}e^{\frac{r+t}{4M}}\\
U & = & \left(1-\frac{r}{2M}\right)^{1/2}e^{\frac{r-t}{4M}}
\end{eqnarray*}
inside the horizon and 
\begin{eqnarray*}
V & = & \left(\frac{r}{2M}-1\right)^{1/2}e^{\frac{r+t}{4M}}\\
U & = & -\left(\frac{r}{2M}-1\right)^{1/2}e^{\frac{r-t}{4M}}
\end{eqnarray*}
outside the horizon. In these coordinates, the future event horizon
is at $U=0$ and the curvature singularity on the hyperboloid $UV=1$.
Time translations in Schwarzschild time act as opposite rescalings
of $U$ and $V$. 

\begin{figure}
\includegraphics[scale=0.5]{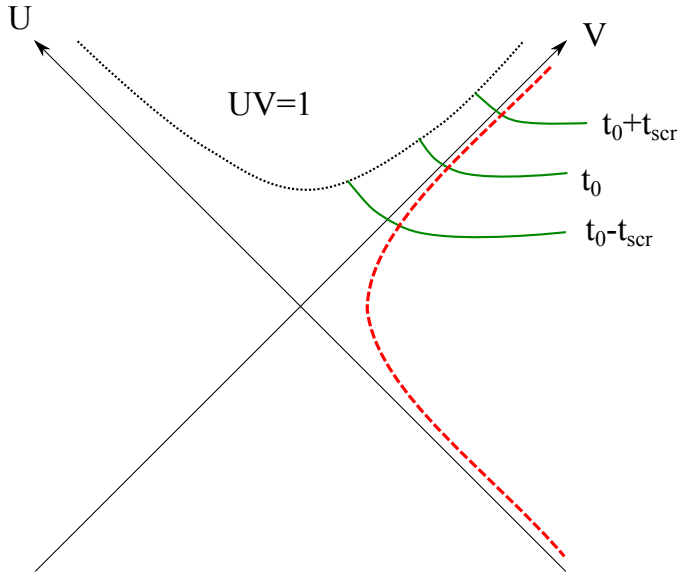}\caption{Schematic figure of time slices labelled by Schwarzschild time $t$
outside the stretched horizon and which approach light sheets inside
the black hole.}

\end{figure}

According to the second postulate of \citep{Susskind:1993if}, physics
outside the so-called stretched horizon is well described by a local
effective field theory, which we'll take to have a UV cutoff $\Lambda$.
The stretched horizon is a timelike surface just outside the event
horizon, located where fiducial observers at rest with respect to
the black hole would measure a local temperature of order the cutoff
scale. In Kruskal coordinates this corresponds to a hyperboloid $UV=-a^{2}$,
where $a$ is a cutoff dependent constant $a\sim(M\Lambda)^{-1/2}$.
The effective field theory of the second postulate is only valid outside
the stretched horizon and is intended for describing observations
made by outside observers. For unitary black hole evolution, it needs
to be supplemented by non-trivial quantum dynamics on the stretched
horizon that serves to absorb, thermalize and re-emit the information
in infalling matter. This outside effective field theory is not well
suited for modeling observations made by infalling observers who enter
the black hole, since, in this description, no reference is made to
the interior geometry of the black hole. Below, we provide an alternative
low-energy effective description, suitable for typical infalling observers,
\emph{i.e. }ones who do not carry with them detailed information about
the quantum state of the black hole. We refer to the Hamiltonian of
the outside effective field theory plus stretched horizon dynamics
as the exact Hamiltonian as it generates the exact S-matrix between
the initial and final states of the system. 

In order to describe infalling observers, we need to introduce a foliation
of the spacetime that covers the black hole interior. Following \citep{Susskind:2012uw},
we adopt a set of time-slices, labelled by Schwarzschild time $t$,
that enter the region inside the horizon of the black hole as shown
in figure 1. Far outside the black hole the time-slices follow the
usual Schwarzschild coordinate system but within a distance of order
$M$ from the stretched horizon the slices turn over and join smoothly
onto surfaces of constant $V$ inside the stretched horizon.

Consider an observer on the $t=t_{0}$ time-slice, who enters the
black hole in radial free fall at $V=V_{0}\gg1$. At the event horizon
the equation for the corresponding radial geodesic simplifies to 
\[
\frac{dU}{d\tau}=\frac{\alpha}{4MV_{0}},\quad\frac{dV}{d\tau}=\frac{eV_{0}}{4M\alpha},
\]
where where $\text{\ensuremath{\alpha}>0}$ parametrizes the instantaneous
velocity and low energy corresponds to $\alpha\sim O(1)$. The worldline
is timelike so $dU/d\tau>0$ everywhere inside the black hole. Assuming
the observer stays in free fall for at least a one Planck unit of
proper time after passing through the horizon, but allowing for arbitrary
timelike motion after that, it follows that the worldline will intersect
the singularity at Kruskal retarded time $U>\frac{\alpha}{4MV_{0}}$.
This in turn implies an upper bound on the advanced Kruskal time when
the observer runs into the singularity given by $V<\frac{4M}{\alpha}V_{0}$. 

Now consider a signal sent into the black hole at Schwarzschild time
$t_{0}+t_{scr}$. The advanced Kruskal time at the point, where the
signal passes through the event horizon, satisfies $V=e^{\frac{t_{scr}}{4M}}V_{0}$
and only the region in the forward light-cone of this point on the
horizon can be influenced by the signal. Therefore, we see that as
long as $ $
\[
t_{scr}>4M\log\frac{4M}{\alpha}
\]
the interior observer will have hit the singularity before the signal
can have any influence. Now if the observer enters the horizon with
a large velocity, this time can be made very long. However, in that
case the energy of the observer in the frame of the black hole is
at least $M_{obs}/\alpha$ if the rest-mass of the observer is $M_{obs}$.
If we demand the back-reaction on the black hole geometry be negligible,
we require
\[
M_{obs}/\alpha\ll M
\]
and as long as 
\[
t_{scr}>8M\log2M
\]
an observer subject to our conditions will always have hit the singularity
prior to receiving the signal. We note this time has the same form
as the fast scrambling time of \citep{Hayden:2007cs}, explaining
our use of the subscript on $t_{scr}$.

\section{Pull-back, push-forward\label{sec:Pull-back,-push-forward}}

The pull-back/push-forward procedure considered in \citep{Susskind:2012uw,Freivogel:2004rd}
gives a prescription for computing correlators of local operators
on a time slice that extends into the black hole interior starting
from data on a late time slice when the black hole has evaporated
and the system only contains outgoing Hawking radiation. The first
step is to use the S-matrix to pull back to a smooth initial state
on an early time slice before the black hole is formed. This state
is then evolved forward using the usual low energy effective field
theory on the time slices of the previous section. An alternate description,
at least for exterior local operators, is provided by evolution with
respect to the exact exterior Hamiltonian. 

An advantage of this approach is that it can be reformulated when
a holographic description of the black hole evaporation is available.
The exterior local Hamiltonian density is a local operator that may
be reconstructed holographically, as can any other local bulk operator,
along the lines of \citep{Kabat:2011rz} (for recent work on the holographic
reconstruction of bulk observables see \citep{Papadodimas:2013anh,Papadodimas:2013jku}).
Thus the two distinct time evolutions, one with respect to the exact
Hamiltonian, and one with respect to the local effective Hamiltonian,
are in principle well-defined.

After a Page time, when half the initial entropy of the black hole
has emerged in the Hawking radiation, the two approaches disagree
when one considers correlators that probe large numbers of outgoing
Hawking particles. In \citep{Susskind:2012uw}, this disagreement
was viewed as supportive of the firewall idea. Our construction gets
around this problem by restricting the pull-back/push-forward prescription
to a finite time interval before and after the infalling observer
enters the black hole.

\section{Decoherence and localization\label{sec:Decoherence-and-localization}}

To better quantify the nature of the disagreement between the two
distinct time evolutions it is helpful to consider the decoherence
of the quantum state as the outgoing Hawking particles stream out,
and potentially interact with measuring apparatus of arbitrarily large
size. This idea of decoherence has a long history going back to the
work of Mott \citep{mott}. He asked the question why do alpha-particle
tracks in a cloud chamber appear to be straight lines when they are
emitted from a nuclear decay in an s-wave. By considering the interaction
of the alpha-particle with the atoms in the cloud chamber, he showed
that after essentially a single interaction, a straight line path
was picked out, with other contributions to the wavefunction interfering
destructively. 

In the present situation, we wish to ask how long it will take for
interactions of the Hawking particles to localize themselves with
respect to some environment. We call this timescale the decoherence
time. If left to their own devices, the self-interaction of these
Hawking particles is so small that the timescale will easily be longer
than the lifetime of the black hole. The question whether an observer
propagating will see local quantum mechanics in their freely falling
frame, or something non-local happen as they approach the horizon,
boils down to a question of calculating the minimal timescale with
which local interactions in the exact theory will lead to a decoherence
of the exact state with respect to local interactions in the exterior.

To obtain the minimal timescale that one might achieve in principle,
imagine surrounding the black hole with a set of detectors, close
to the horizon. Such a set of detectors will behave much like the
stretched horizon itself. Specifically, we seek the timescale with
which an incoming state hitting the stretched horizon should subsequently
decohere due to local interactions of the emitted Hawking particles
with the detectors. Since the entanglement is not emitted until after
the scrambling time \citep{Hayden:2007cs}, we expect the timescale
for decoherence will be bounded below by $t_{scr}$ (with respect
to the timeslices of section \ref{sec:Black-hole-geometry}).

If we apply this picture to the attempt at reconstructing the black
hole interior in section \ref{sec:Pull-back,-push-forward} we immediately
see a problem. The Page time is much longer than this decoherence
time. Already after $t_{scr}$ the state will effectively decohere
due to the local interactions of the exterior Hawking particles with
potentially large, localized detectors outside the black hole. Such
interactions will appear highly non-local from the viewpoint of the
interior effective description. Thus interior observers will not see
ordinary quantum evolution with respect to their local Hamiltonian
density.

\section{Pull-back/push forward revisited\label{sec:Pull-Back/Push-Forward-Revisited}}

Let us instead try to introduce the minimal elements needed to build
an interior description of the black hole from the point of view of
some set of observers close to some pencil of timelike geodesics that
cross the horizon. Let such an observer cross the horizon at $t_{0}$,
following the discussion of section \ref{sec:Black-hole-geometry}
where the timeslices of interest are set up. The decoherence arguments
of section \ref{sec:Decoherence-and-localization} indicate that at
best we can trust evolution with respect to the local effective Hamiltonian
in a time interval $t_{0}-t_{scr}<t<t_{0}+t_{scr}$. 

We set up the local effective field theory description of this restricted
class of infalling observers using a version of the pull-back/push-forward
procedure as follows. We use the exact Hamiltonian, including stretched
horizon degrees of freedom, to evolve to the timeslice $t_{0}-t_{scr}$.
This specifies the initial state outside the stretched horizon, but
in order to follow the observer into the black hole we must further
specify the initial state inside the stretched horizon. The arguments
of section \ref{sec:Black-hole-geometry} show that with a reasonable
proper distance cutoff, the details of the initial state at $t_{0}-t_{scr}$
in the interior are irrelevant once one propagates forward to $t_{0}$
for all but a thin layer extending from of order a Planck length inside
the global horizon to the stretched horizon. 

To specify this remaining initial data at $t_{0}-t_{scr}$, we place
vacuum initial conditions in this layer. These initial conditions
should be determined by the condition that the state be a good approximation
to a Hadamard state \citep{Kay:1988mu,Radzikowski:1996pa}. It should
be noted that such a state leads to a firewall inside the global horizon,
as originally suggested in \citep{Almheiri:2012rt}. The condition
of a Hadamard state means that the local energy density will be relatively
small in the thin layer. Likewise, in the exterior, the arguments
of \citep{Lowe:2013zxa} show that the expectation value of the stress
tensor seen by a freely falling observer will be very close to the
 result expected in the Hartle-Hawking or Unruh vacua. If one also
introduces a Planck scale smearing in the spatial directions, the
computation of \citep{Lowe:2013zxa} shows the correction to the energy
density expected, beyond the purely thermal results, will be of order
$e^{-S(M)}$ in Planck units, where $S(M)$ is the Bekenstein-Hawking
entropy. However as one leaves the layer, moving inward, one encounters
modes that are not entangled with their exterior partners, as they
would be in the Unruh or Hartle-Hawking vacua, so one expects an energy
density there corresponding to an effective temperature of order the
stretched horizon cutoff scale. 

The beauty of the construction is that the geometry described in section
\ref{sec:Black-hole-geometry} is such that this interior firewall
will hit the singularity before it can interact with our observer
entering at $t_{0}$. Taking this initial state at $t_{0}-t_{scr}$
and pushing forward to $t_{0}$ using the effective local Hamiltonian
then leads to a good initial state at $t_{0}$ for the infalling observer.
In particular, it solves the so-called frozen vacuum problem \citep{Bousso:2013ifa},
because the only infalling data that can influence the infalling observer,
inside the horizon, falls in later than $t_{0}-t_{scr}$ by section
\ref{sec:Black-hole-geometry}. Such data will interact and change
the state in the interior layer as one evolves forward to $t_{0}$,
by which time we will typically have a non-vacuum initial state in
the interior.

The need to specify vacuum initial data in this Planck layer renders
the construction of the interior effective field theory non-unitary,
which is a key difference from the assumptions of \citep{Almheiri:2012rt}.
However, as noted above, this is reasonable, because the infaller
can only effectively come into causal contact with signals that entered
the black hole within a scrambling time. The situation is similar
to setting up effective field theory in an expanding patch of the
de Sitter spacetime, with a fixed proper distance ultraviolet cutoff.
There vacuum modes are added by hand as the patch expands, which again
is formally a non-unitary process, unless a strict continuum limit
can be defined. This addition of short distance modes reconciles the
arbitrarily large number of short distance degrees of freedom in the
interior effective field theory with the finite number of black hole
states at fixed energy of the exact description. The overabundance
of interior degrees of freedom is a necessary artifact of the field
theory description, but due to the limited measurement precision available
to an interior observer ensures this does not lead to contradiction,
as emphasized in \citep{Lowe:2006xm}.

The recipe described above thus gives a regular time evolution for
the interior observer until near the curvature singularity. This local
evolution of the interior observer has a non-local interpretation
in the exterior stretched horizon theory prior to $t_{0}+t_{scr}$,
that only comes into conflict with the subsequent emission of Hawking
radiation after the time $t_{0}+t_{scr}$, as was argued in section
\ref{sec:Decoherence-and-localization}. By this time, however, the
observer has already hit the singularity by the arguments of section
\ref{sec:Black-hole-geometry}. For a finely tuned external state,
as might be arranged by some large external measuring device, time
evolution may lead to an ingoing state entangled with a Hawking particle
emerging from the stretched horizon, just as the observer crosses.
Such a state will show up as a kind of fireball for the observer.
If the argument of typicality of black hole states of \citep{Avery:2013vwa}
is correct, then such fireballs will quickly evolve back to a smooth
apparent geometry. The same kind of finely tuned firewall may also
be arranged to appear inside the horizon. In this case the entangled
pair of modes is inside a future trapped region so both modes will
be ingoing. It has been suggested that this kind of fine tuning may
require manipulations of the external measurement apparatus that cannot
be carried out within the black hole lifetime \citep{Harlow:2013tf},
however the present construction only requires this cannot be done
faster than $t_{scr}$.

It should be noted that our recipe will only work for typical observers
who are not able to measure correlators of a large number of local
operators, or resolve differences of order $e^{-S(M)}$ in correlators
of small numbers of local operators, since the arguments of \citep{Lowe:2013zxa}
are used. The timeslice at $t_{0}$ is certainly capable of accommodating
large measuring machines, that are not necessarily subject to these
restrictions. Correlators of local observables will agree between
the low-energy effective description and the exact exterior description
in the overlap region outside the stretched horizon between $t_{0}-t_{scr}$
and $t_{0}+t_{scr}$, unless the local operators are somehow able
to probe what is usually nonlocal entanglement between the Hawking
particles emitted from the stretched horizon after $t_{0}-t_{scr}$
and those emitted earlier. 

Restrictions on the measurements of such typical observers have also
been studied in the context of fuzzball complementarity \citep{Mathur:2012jk,Mathur:2013gua}
and the need for a sequence of patches of effective field theories
to describe the quantum mechanics of an inside observer was noted
in \citep{Mathur:2013talk}. Our effective field theory approach nevertheless
differs from fuzzball complementarity in that it gives a detailed
construction that provides an approximate interior description that
might be realized in any unitary model of black hole evaporation.

\section{Conclusions\label{sec:Conclusions}}

We have presented an approximate effective field theory to model observations
made by a typical low-energy observer entering a black hole in free
fall at a prescribed time. The effective field theory is allowed to
be only approximate because the measurement precision that is available
to such an observer is limited both by the finite proper time remaining
before hitting the singularity and by the finite size of measuring
devices that can be carried into the black hole without significant
back-reaction on the geometry \citep{Lowe:2006xm}. Our construction
involves a variant of a pull-back/push-forward procedure that takes
into account the minimal decoherence time scale of outgoing Hawking
quanta and only operates within a relatively short time interval before
and after the infalling observer enters the black hole. The specification
of the initial data involves a mildly non-unitary step, amounting
to putting short distance modes in their vacuum state as they emerge
below a short distance ultraviolet cutoff. This is a necessary artifact
of the interior field theory representation of the physics, but does
not change the unitary exterior description.

We argue that a typical observer inside a typical black hole will
see no quantum drama until they approach the singularity. On the other
hand, an external influence, having acquired precise knowledge of
the black hole initial state, is capable of sending in a low energy
ingoing component of the state, precisely entangled with some outgoing
Hawking particle. While such a process requires extreme fine-tuning,
it would cause our recipe for the ``inside view'' to fail for some
particular infalling observer who encounters the resulting firewall.
Such a failure is an inevitable consequence of the approximate description
of the interior extracted from the exact evolution, and we believe
in this case the exception proves the rule.

\paragraph*{Acknowledgements}

L.T. thanks Daniel Harlow and I-Sheng Yang for useful discussions.
The research of D.A.L. is supported in part by DOE grant DE-SC0010010-Task
A and an FQXi grant. The research of L.T. is supported in part by
grants from the Icelandic Research Fund grant 130131-052 and the University
of Iceland Research Fund.

\bibliographystyle{elsarticle-num}
\bibliography{firewall2}

\end{document}